\documentclass[runningheads]{llncs}

\usepackage[T1]{fontenc}
\usepackage[utf8]{inputenc}
\usepackage{graphicx}
\usepackage{booktabs}
\usepackage{array}
\usepackage{multirow}
\usepackage{amsmath}
\usepackage{amssymb}
\usepackage{xcolor}
\usepackage{rotating}      
\usepackage{siunitx}
\usepackage{caption}
\usepackage[hidelinks]{hyperref}

\newcommand{\best}[1]{\textbf{#1}}   

\begin{document}

\title{A Comprehensive Objective Evaluation of Modern Text-to-Speech for Turkish\\
Using Speech Quality Assessment Models}
\titlerunning{Objective Evaluation of Turkish TTS}
\author{Yunus Emre Ozkose \and
Alperen Kahraman \and
Ali Haznedaroglu}
\authorrunning{Y. E. Ozkose et al.}
\institute{Sestek, Ankara, T\"urkiye\\
\email{\{yunusemre.ozkose, alperen.kahraman, ali.haznedaroglu\}@sestek.com}}

\maketitle

\begin{abstract}
Modern text-to-speech (TTS) systems can clone a target speaker from a short reference clip or be fine-tuned on a target voice, yet their behaviour on morphologically rich, lower-resource languages such as Turkish remain under-characterised. We present a systematic benchmark of four contemporary systems (Chatterbox, CosyVoice, OmniVoice, and VoxCPM2) evaluated across fine-tuned and zero-shot configurations, contrasted with a conventional VITS baseline and anchored to natural gold speech. Each configuration is scored with eighteen complementary objective metrics spanning learned naturalness predictors (UTMOS\,v2, DNSMOS-Pro, SCOREQ, WhisQA, AudioBox-PQ, NatScore, SpeechLMScore), intelligibility and signal-quality estimators (SQUIM PESQ/SI-SDR/STOI, Brouhaha), speaker similarity, distributional fidelity (TTSDS) and low-level acoustic descriptors. We further analyse how quality varies with utterance length and quantify long-form temporal consistency through speaker-identity and naturalness drift over chunked utterances. We release our evaluation code to support reproducible TTS evaluation.

\end{abstract}

\section{Introduction}
\label{sec:intro}

Speech synthesis is one of the most important tasks for various speech-oriented pipelines such as voice chatbots, virtual assistants, etc ... The primary objective is to generate a qualified synthesis from the given text. While it is a crucial part of various pipelines, the evaluation and selection of models are still hard and time-consuming because of metrics such as mean opinion score (MOS) and AB test, which require human investigation. Several works try to mimic MOS score, such as UTMOS~\cite{saeki22c_interspeech,baba2024utmosv2}. However, observing MOS score is not enough to select a text-to-speech (TTS) model since speech includes multiple quality dimensions like naturalness, signal quality, prominence, etc. Hence, a more comprehensive strategy is need to select metrics and methods.

In recent years, TTS models have achieved high success in natural and high-quality synthesis. Models have moved from per-speaker training towards zero-shot synthesis, where a single model is able to produce an unseen speaker synthesis from only a few reference audio without speaker-based finetuning. The latest systems~\cite{cosyvoice3,gong2026mossttstechnicalreport} clone timbre, prosody, and speaking style from a short prompt, and bring a high-quality voice cloning without the high cost of speaker-specific data collection. Almost all of these evaluations and achievements are demonstrated on relatively high-resource languages such as English and Chinese. Turkish is comparatively underserved where it is an agglutinative language with vowel harmony, productive suffixation, and a largely transparent orthography. To our knowledge, no prior study evaluates several state-of-the-art zero-shot systems on Turkish under a single, controlled, reference-aware protocol.



In this paper, we benchmark four state-of-the-art TTS models which span complementary design philosophies (LLM-plus-flow-matching, discrete diffusion-language-model
decoding and tokenizer-free continuous diffusion) in various states (finetuned, zero-shot, text-based, phoneme-based). We contrast them with a conventional, non-zero-shot VITS baseline. These models are evaluated under eighteen objective metrics covering naturalness,
intelligibility, signal quality, speaker similarity, distributional fidelity, and
low-level acoustics. Beyond these comparisons, we investigate whether models are able to generate consistent synthesis or not. Our analysis yields several findings that several automatic speech assessment methods are need to be observed for specific demand, and an averaged leaderboard would obscure. A single system cannot win by itself, but per-metric winners partition cleanly by metric family. Several models even surpass the gold reference on signal-clean predictors, exposing a tension between metrics that reward clean, intelligible audio and those that track human-likeness. Speaker similarity, the one dimension on which natural speech reliably tops every system, emerges as the most trustworthy single indicator of cloning fidelity.

In summary, our contributions are: (i) a reproducible, multi-metric benchmark of TTS models in Turkish, covering four contemporary systems and a conventional baseline against a natural-speech anchor across eighteen objective metrics; (ii) an analysis showing that relative system
rankings are largely determined by the chosen metric family and by utterance length,
together with evidence that several signal-oriented predictors reward idealised audio
over genuine naturalness; and (iii) a long-form temporal-consistency study that
separately quantifies speaker-identity drift and naturalness drift over an utterance. Public release of our evaluation code.\footnote{Code and results: \url{https://github.com/EmreOzkose/tr-tts-eval}}

\section{Related Work}
\label{sec:related}

\subsection{Modern Speech Synthesis}
Recent TTS models have shifted from traditional parametric or purely end-to-end neural architectures toward generative, zero-shot voice cloning systems. While more traditional architectures like VITS~\cite{vits} successfully integrated acoustic and waveform generation into a unified, high-quality framework, they remain constrained by the requirement of per-speaker training and struggle with unseen target voices. Modern TTS methods~\cite{omnivoice,cosyvoice3} address this limitation by reframing speech synthesis as a conditional generation task, where a speaker's identity is cloned directly from a short, unannotated reference audio clip without any fine-tuning. The neural codec language model
line VALLE~\cite{valle} treats TTS as language modelling over discrete
audio codes and clones a voice from a few seconds prompt while flow and
diffusion based systems such as NaturalSpeech2~\cite{naturalspeech2} pursued the
same goal through continuous representations. Multilingual voice-cloning systems, including YourTTS~\cite{yourtts} and XTTS~\cite{xtts}, extended zero-shot synthesis to many languages, often covering low and medium resource ones. The four systems evaluated here continue this trajectory along complementary design axes: CosyVoice~\cite{cosyvoice3} couples an LLM over supervised semantic tokens with flow matching; OmniVoice~\cite{omnivoice} uses a diffusion-language-model-style discrete non-autoregressive decoder over multi-codebook acoustic tokens; VoxCPM2~\cite{voxcpm2} dispenses with discrete tokens entirely, modelling a continuous acoustic space with a diffusion-autoregressive pipeline; and Chatterbox~\cite{chatterboxtts2025} pairs an LLM-style token model with a flow-matching decoder and emotion control. We include the conventional, non-zero-shot VITS~\cite{vits} as a well-understood reference point. Despite this rich landscape, head-to-head comparisons on Turkish are scarce, which is the gap this paper addresses.

\subsection{Evaluation of Synthetic Speech}
Automatic speech quality assessment has seen significant advancement with the advent of deep-learning-based models designed to mimic human Mean Opinion Scores (MOS) since subjective listening tests are slow and costly. This has driven a large body of work on automatic predictors: learned naturalness/MOS estimators such as UTMOS~\cite{baba2024utmosv2} and DNSMOS~\cite{reddy2021dnsmos}; reference-free intelligibility and signal-quality measures such as TorchAudio-SQUIM~\cite{torchaudio-squim} (PESQ/SI-SDR/STOI) and SCOREQ~\cite{ragano2024scoreq}; acoustic-condition estimators such as Brouhaha~\cite{lavechin2023brouhaha}; and distribution-level scores such as TTSDS~\cite{minixhofer2025ttsds2}, which compare the feature distribution of synthetic and natural speech. Each predictor captures only a slice of perceptual quality and can disagree with the others.

Classic subjective speech quality assessment models also suffer from a lack of robustness and poor out-of-domain generalization when applied to unseen acoustic environments or speech synthesis systems. To address these systemic validation gaps, Huang et al.~\cite{huang2026mos} introduced MOS-Bench, a comprehensive evaluation benchmark containing diverse training and testing partitions specifically curated to evaluate the out-of-distribution generalization capabilities of existing metrics. Recognizing that single-score absolute regression struggles with inconsistent rating standards across heterogeneous datasets, recent efforts have shifted toward preference-based or comparative frameworks. For instance, Cao et al.~\cite{cao2025scores} formulated MOS-RMBench, a benchmark that reframes historical MOS datasets into a preference-comparison setting tailored for speech quality reward modeling, demonstrating that generative reward models utilizing a MOS-difference-based reward function can markedly improve fine-grained quality discrimination. Building upon these paradigms, UrgentMOS~\cite{wang2026urgentmos} bridges absolute and relative paradigms by proposing a unified framework that jointly learns from multi-metric objective and perceptual targets while explicitly modeling pairwise quality preferences via comparative MOS (CMOS) prediction.

\section{Systems Under Evaluation}
\label{sec:systems}

We conduct experiments with four modern TTS models and one conventional model. Natural speech (\emph{gold}) records serve as a gold reference anchor. All systems are open-source and capable to generate a specific speaker's speech. They have different design philosophies: LLM-plus-flow-matching, tokenizer-free continuous diffusion, and diffusion-language-model discrete NAR, which makes them a representative cross-section of the current state of the art.

\begin{description}
  \item[CosyVoice~\cite{cosyvoice3} (finetuned)] A scalable multilingual zero-shot synthesizer from the Alibaba FunAudioLLM group. It represents speech with supervised semantic tokens derived from a multilingual ASR encoder, generates these tokens with an LLM, and reconstructs the waveform with a conditional flow-matching model conditioned on a speaker embedding. We evaluate it under two textual front-ends: \emph{raw text} and \emph{phonemes} to isolate the effect of the input representation on a morphologically rich language such as Turkish.
  \item[OmniVoice~\cite{omnivoice} (finetuned and zero-shot)] A massively multilingual zero-shot model from Xiaomi that scales to $600+$ languages. Rather than a two-stage text-to-semantic-to-acoustic pipeline, it uses a diffusion-language-model style discrete non-autoregressive architecture that maps text directly to multi-codebook acoustic tokens, trained with full-codebook random masking and initialised from a pre-trained LLM for intelligibility. We evaluate two configurations: a \emph{finetuned} mode and a \emph{zero-shot} mode.
  \item[Chatterbox~\cite{chatterboxtts2025} (zero-shot)] An open-source model from Resemble AI built on a $0.5$\,B-parameter Llama-style backbone and trained on roughly $500$k hours of speech. It couples a text-to-token model with a flow-matching token-to-waveform decoder and supports zero-shot cloning from a short reference clip. We use its multilingual variant, which extends the system to Turkish among $23$ languages.
  \item[VoxCPM2~\cite{voxcpm2} (zero-shot)] A tokenizer-free system from OpenBMB built on a MiniCPM-4 backbone. The evaluated checkpoint is the multilingual VoxCPM2 release.
  \item[VITS~\cite{vits} (finetuned)] A conventional, non-zero-shot end-to-end model that combines a conditional variational autoencoder with normalising flows, adversarial training, and a stochastic duration predictor. It is \emph{not} a voice-cloning system and is included as a strong, well-understood reference point against which the zero-shot systems are contrasted.
\end{description}

Chatterbox and VoxCPM2 are evaluated for only the zero-shot case. VITS is fine-tuned for a specific speaker who has 20 hours of records. Cosyvoice and Omnivoice are finetuned with 2000 hours of speech, which includes 4 languages (English, Arabic, Turkish, and Urdu) and 22 speakers. This setup also includes 20 hours of speaker speech, which is in VITS training. All zero-shot cases are using this specific speaker's 9-second prompt:

\begin{quote}
    \textit{Bir serin rüzgarın dokunuşundan sonra keşfettiği ateşe, başının, ciğerlerinin, bütün varlığının ateşine bu üşümenin şifa vereceğini zannediyordu.}
\end{quote}

\section{Evaluation Methodology}
\label{sec:method}

We conduct experiments with various objective metrics on synthesised audio and on natural
reference audio. Metrics are grouped by the perceptual dimension they target.

\subsection{Reference-Free Naturalness and Quality Predictors}
UTMOSv2\cite{baba2024utmosv2}, DNSMOS-Pro~\cite{reddy2021dnsmos} (with its BVCC, NISQA, and VCC2018 heads), SCOREQ~\cite{ragano2024scoreq}(\textit{synthetic-nr}), WhisQA~\cite{close2025whisqa}, AudioBox-PQ~\cite{Audiobox-aes}, NatScore~\cite{singh2026natscore}, and SpeechLMScore~\cite{maiti2023speechlmscore} are learned predictors of perceived naturalness and overall quality.

\subsection{Intelligibility and Signal-Level Metrics}
SQUIM (PESQ, SDR, and STOI) ~\cite{torchaudio-squim} provides reference-free estimates of perceptual quality, signal-to-distortion ratio, and intelligibility, respectively. Brouhaha~\cite{lavechin2023brouhaha} estimates speech/noise characteristics (reported here as an SNR-like score).

\subsection{Speaker Similarity}
Speaker similarity is the cosine similarity between speaker embeddings of the synthesised utterance and the target reference; values approach the gold anchor as cloning fidelity improves. Wespeaker~\cite{wespeaker} is used to compute speaker embeddings. We use ResNet221-LM speaker verification model that is trained on VoxCeleb~\cite{nagrani2017voxceleb}

\subsection{Distributional Fidelity}
TTSDS (TTS Distribution Score) v2~\cite{minixhofer2025ttsds2} measures how closely the distribution of synthetic speech features matches that of natural speech.

\subsection{Low-Level Acoustic Descriptors}
Integrated loudness (LUFS) and a sibilance ratio characterise low-level acoustic
behaviour; for these, proximity to the gold reference is the relevant criterion.

\subsection{Long-Form Temporal Consistency}
\label{sec:method-temporal}
To probe stability over long utterances, each utterance is split into chunks, and 2 metrics (speaker similarity and UTMOS scores) are tracked across chunks. We report average similarity to the reference (\textit{mean\_sim\_vs\_ref}), std of similarity across chunks; lower is
more stable (\textit{consistency\_std}), linear trend of similarity over the utterance (\textit{drift\_slope}), and similarity change from first to last chunk (\textit{first\_last\_delta}).


\section{Experimental Setup}
\label{sec:setup}

For evaluation, a small-scale test dataset was constructed. A subset was sampled from the test set of the existing 20-hour Turkish speaker dataset, consisting of 100 short, 100 medium-length, and 100 long utterances selected at random. Short, medium-length, and long utterances were defined based on character length intervals of 0–50, 51–150, and greater than 150 characters, respectively. The 20-hour speaker dataset is a TTS dataset that we collect internally to train our TTS models. Sampling rate is 48 kHz originally, but we adapt samples according to model requirements.

\section{Results}
\label{sec:results}

\subsection{Overall Comparison}
\label{sec:results-overall}

\begin{figure}[t]
\centering
\includegraphics[width=0.82\textwidth]{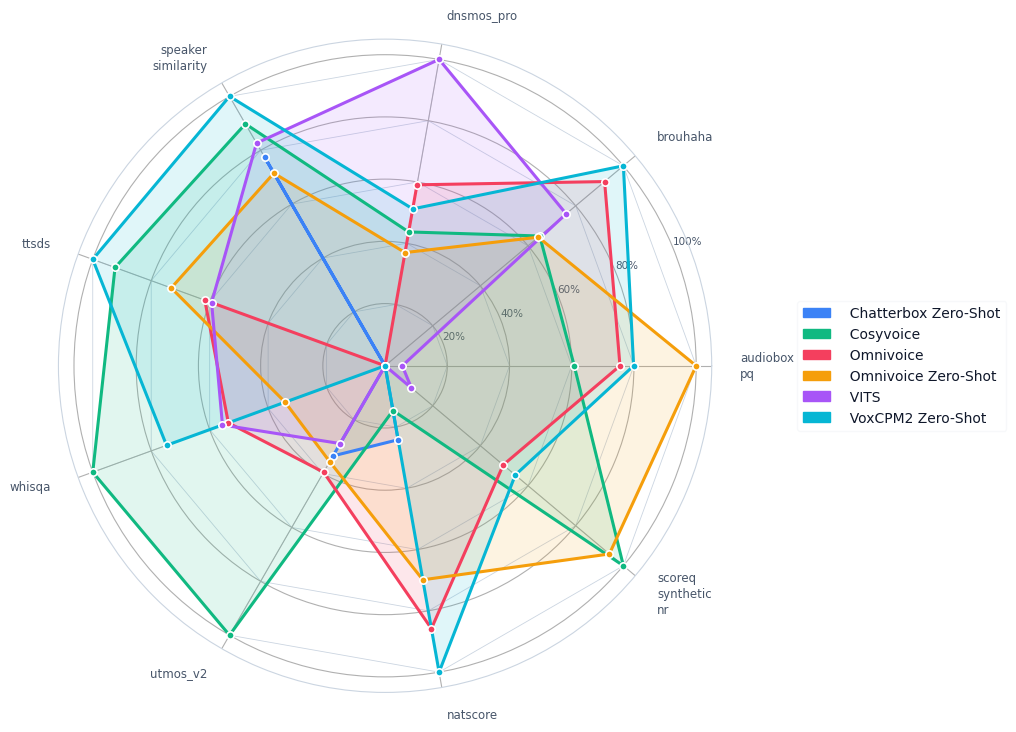}
\caption{Radar comparison of metrics, normalized to a common
$[0,100]\%$ scale, with CosyVoice represented by its canonical raw-text
configuration. Legend annotations report, per system, the number of headlines
}
\label{fig:radar}
\end{figure}

All averaged metric results over the full test set for all models and the gold set are reported in Table~\ref{tab:overall}. For each metric, the best result is shown in \best{bold}. The gold column is an anchor, not a competitor. Figure~\ref{fig:radar} summarises the nine headline metrics on a common normalized scale and annotates, per system, the number of headline metrics it wins.

\begin{table}[t]
\centering
\caption{Overall objective metrics (mean over the full test set). Best synthetic
system per row in \best{bold}. CosyVoice is represented by its canonical raw-text
configuration. Columns:
CB = Chatterbox (zero-shot),
CV = CosyVoice (raw text),
OV = OmniVoice (finetuned),
OV-Z = OmniVoice (zero-shot),
VITS = VITS,
VOX = VoxCPM2 (zero-shot),
Gold = natural reference.}
\label{tab:overall}
\setlength{\tabcolsep}{4pt}
\resizebox{\textwidth}{!}{%
\begin{tabular}{l |rrrrrr| r}
\toprule
Metric & CB & CV & OV & OV-Z & VITS & VOX & Gold \\
\midrule
audiobox-pq            & 7.643 & 7.793 & 7.785 & \best{7.831} & 7.654 & 7.793 & 7.657 \\
brouhaha                & 50.132 & 60.249 & 61.498 & 58.066 & 59.507 & \best{62.471} & 55.309 \\
dnsmos\_pro             & 3.564 & 3.770 & 3.954 & 3.808 & \best{4.225} & 3.902 & 3.601 \\
dnsmos-pro-bvcc       & 3.347 & 3.643 & \best{3.745} & 3.522 & 3.616 & 3.533 & 3.698 \\
dnsmos-pro-nisqa      & 3.564 & 3.770 & 3.954 & 3.808 & \best{4.225} & 3.902 & 3.601 \\
dnsmos-pro-vcc2018    & 3.045 & 2.925 & 2.722 & 2.773 & \best{3.138} & 2.886 & 3.171 \\
speaker-similarity     & 0.873 & 0.878 & 0.843 & 0.870 & 0.875 & \best{0.881} & 0.894 \\
squim-pesq             & 0.995 & 0.993 & 0.996 & \best{0.996} & 0.992 & 0.995 & 0.992 \\
squim-si-sdr          & 24.939 & 25.736 & 27.119 & \best{28.523} & 25.211 & 27.053 & 24.455 \\
squim-stoi             & 3.762 & 3.753 & 3.772 & \best{3.926} & 3.690 & 3.788 & 3.639 \\
utmos-v2               & 3.092 & \best{3.441} & 3.129 & 3.105 & 3.064 & 2.886 & 3.079 \\
low-level-lufs        & -26.624 & \best{-22.443} & -24.163 & -27.600 & -23.989 & -25.601 & -25.151 \\
low-level-sib.-ratio & 0.041 & 0.050 & 0.068 & 0.070 & 0.053 & \best{0.077} & 0.084 \\
scoreq-synthetic-nr   & 3.697 & \best{4.063} & 3.888 & 4.059 & 3.740 & 3.907 & 3.782 \\
speechlmscore           & 1.592 & 1.672 & 1.636 & 1.637 & \best{1.733} & 1.623 & 1.703 \\
whisqa                  & 4.341 & \best{4.720} & 4.509 & 4.448 & 4.515 & 4.575 & 4.331 \\
natscore                & -1.447 & -1.233 & -1.016 & -1.128 & -1.616 & \best{-0.917} & -1.384 \\
ttsds                   & 77.012 & \best{79.735} & 78.334 & 78.580 & 78.284 & 79.156 & 77.864 \\
\bottomrule
\end{tabular}}
\end{table}

First of all, \emph{no system wins globally}: Table~\ref{tab:overall} shows that the per-metric best is spread across five different entries that cluster cleanly by family. VITS dominates the \textit{DNSMOS-Pro} family and \textit{speechlmscore}; OmniVoice~(zero-shot) sweeps the SQUIM intelligibility/signal triplet (\textit{squim-pesq}, \textit{squim-si-sdr}, \textit{squim-stoi}) together with \textit{audiobox-pq}; CosyVoice owns the learned-naturalness/distributional cluster (\textit{utmos-v2}, \textit{scoreq-synthetic-nr}, \textit{whisqa}, \textit{ttsds}); and VoxCPM2 leads on speaker similarity, \textit{natscore} and \textit{brouhaha}. Counting per-metric winners over the full battery, these four systems take three to four metrics each, OmniVoice~(raw) takes one, and Chatterbox none.

Secondly, several synthetic systems \emph{exceed} the gold reference on a number of predictors: VITS on \textit{dnsmos\_pro} ($4.225$ vs.\ $3.601$), OmniVoice~ZS on \textit{squim-si-sdr} ($28.5$ vs.\ $24.5$), and \textit{audiobox-pq}, and VoxCPM2 on \textit{natscore} ($-0.92$ vs.\ $-1.38$). 

Because natural speech cannot plausibly be \textit{less natural} than synthesis, these inversions indicate that the signal-quality and clean-MOS predictors reward idealised, low-noise, smoothly voiced audio rather than perceptual realism. The metric on which gold remains clearly highest is \textit{speaker-similarity} ($0.894$); among synthetic systems VoxCPM2 is closest ($0.881$) and OmniVoice~(raw) the furthest ($0.843$). Finally, Chatterbox is the only system that wins no metric. Sec.~\ref{sec:results-length} shows this is driven largely by a collapse on short utterances. For the low-level descriptors, proximity to gold is the relevant reading: among the systems in Table~\ref{tab:overall}, VoxCPM2 comes closest to the gold \textit{low\_level\_sibilance\_ratio} ($0.077$ vs.\ $0.084$), whereas the highlighted \textit{low-level-lufs} cell merely marks the loudest system and is in fact further from the gold loudness than several others.

\subsection{Effect of Utterance Length}
\label{sec:results-length}
Evaluation by length gropus are given in Table~\ref{tab:length}. Results show that the hardest setup is short-length group since for almost every system and metric, scores are lowest in the short group and recover for medium and long utterances. The most dramatic case is Chatterbox, which collapses on short input (\textit{squim-si-sdr}) falls from $26.4$ (long) to $20.4$ (short), speaker similarity from $0.889$ to $0.825$, \textit{whisqa} from $4.58$ to $3.61$, and \textit{natscore} from $-1.28$ to $-2.22$, and its \textit{brouhaha} uniquely \emph{drops} on short input ($53.2\!\rightarrow\!40.5$) while it rises for every other system. By contrast the CosyVoice, OmniVoice and VoxCPM2 configurations remain comparatively length-robust, with OmniVoice~ZS holding the highest \textit{squim-si-sdr} across all three groups and VoxCPM2 the best \textit{natscore} in the long and medium groups (CosyVoice~(raw) edges it on short utterances). Speaker similarity declines with shorter utterances for \emph{every} entry including gold ($0.917\!\rightarrow\!0.827$); since the natural reference degrades in the same direction, part of this drop is an artifact of estimating speaker embeddings from very short audio rather than a synthesis defect.

\begin{table}[t]
\centering
\caption{Quality by utterance-length group for representative metrics. APQ = audiobox-pq, BRH = brouhaha, DM = dnsmos\_pro, SPK = speaker-similarity, SISDR = squim-si-sdr, UTM = utmos-v2, WHQ = whisqa, NAT = natscore.}
\label{tab:length}
\setlength{\tabcolsep}{4pt}
\resizebox{\textwidth}{!}{%
\begin{tabular}{l l rrrrrrrr}
\toprule
Len. & System & APQ & BRH & DM & SPK & SISDR & UTM & WHQ & NAT \\
\midrule
\multirow{7}{*}{long}
 & Chatterbox      & 7.735 & 53.18 & 3.654 & 0.889 & 26.39 & 3.179 & 4.578 & -1.282 \\
 & CosyVoice       & 7.836 & 58.36 & 3.819 & 0.900 & 25.35 & 3.465 & 4.765 & -1.308 \\
 & OmniVoice       & 7.791 & 58.12 & 3.948 & 0.872 & 26.23 & 3.135 & 4.581 & -1.027 \\
 & OmniVoice (Z)   & 7.852 & 55.10 & 3.770 & 0.887 & 28.49 & 3.165 & 4.526 & -1.064 \\
 & VITS            & 7.724 & 56.76 & 4.228 & 0.894 & 24.80 & 3.077 & 4.582 & -1.641 \\
 & VoxCPM2         & 7.820 & 59.23 & 3.926 & 0.899 & 26.85 & 2.914 & 4.640 & -0.803 \\
 & gold            & 7.722 & 53.09 & 3.641 & 0.917 & 24.03 & 3.124 & 4.385 & -1.420 \\
\midrule
\multirow{7}{*}{medium}
 & Chatterbox      & 7.697 & 51.47 & 3.565 & 0.877 & 25.57 & 3.164 & 4.434 & -1.203 \\
 & CosyVoice       & 7.819 & 60.73 & 3.788 & 0.883 & 26.20 & 3.461 & 4.759 & -1.090 \\
 & OmniVoice       & 7.828 & 63.70 & 4.016 & 0.850 & 27.94 & 3.168 & 4.591 & -0.779 \\
 & OmniVoice (Z)   & 7.851 & 58.79 & 3.891 & 0.877 & 28.58 & 3.213 & 4.539 & -0.864 \\
 & VITS            & 7.686 & 60.46 & 4.269 & 0.881 & 25.33 & 3.117 & 4.579 & -1.562 \\
 & VoxCPM2         & 7.819 & 64.11 & 3.928 & 0.887 & 27.50 & 2.912 & 4.646 & -0.677 \\
 & gold            & 7.705 & 56.21 & 3.662 & 0.901 & 24.28 & 3.116 & 4.422 & -1.241 \\
\midrule
\multirow{7}{*}{short}
 & Chatterbox      & 7.332 & 40.49 & 3.335 & 0.825 & 20.36 & 2.766 & 3.607 & -2.221 \\
 & CosyVoice       & 7.647 & 64.33 & 3.621 & 0.814 & 26.03 & 3.350 & 4.546 & -1.250 \\
 & OmniVoice       & 7.703 & 66.81 & 3.879 & 0.757 & 28.18 & 3.057 & 4.208 & -1.339 \\
 & OmniVoice (Z)   & 7.746 & 64.53 & 3.784 & 0.818 & 28.54 & 2.927 & 4.118 & -1.675 \\
 & VITS            & 7.426 & 65.06 & 4.151 & 0.817 & 26.07 & 2.949 & 4.253 & -1.631 \\
 & VoxCPM2         & 7.685 & 68.27 & 3.805 & 0.830 & 26.91 & 2.776 & 4.307 & -1.556 \\
 & gold            & 7.420 & 59.61 & 3.412 & 0.827 & 25.80 & 2.909 & 4.058 & -1.501 \\
\bottomrule
\end{tabular}}
\end{table}

\subsection{Long-Form Temporal Consistency}
\label{sec:results-temporal}

\begin{figure}[t]
\centering
\includegraphics[width=0.92\textwidth]{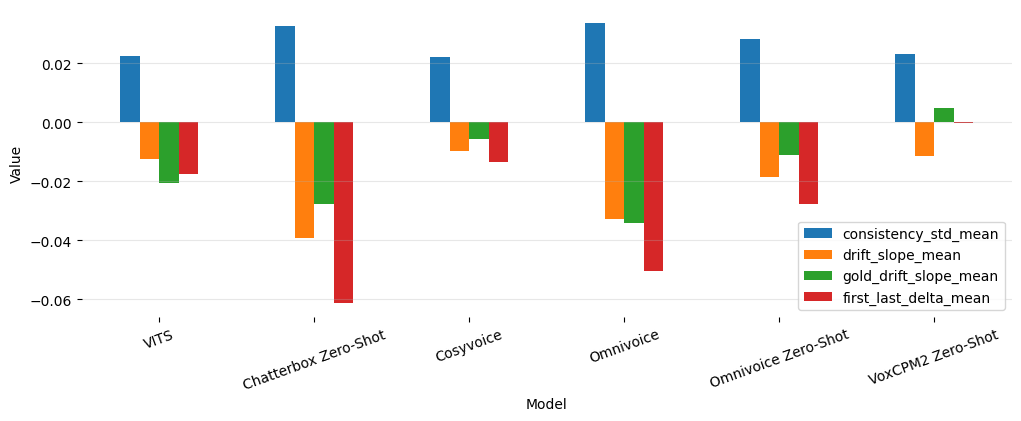}
\caption{Drift and consistency metrics per system: \textit{consistency\_std\_mean},
\textit{drift\_slope\_mean}, \textit{gold\_drift\_slope\_mean}, \textit{first\_last\_delta\_mean}.}
\label{fig:temporal-drift}
\end{figure}

To investigate consistency over time in synthetic speech, we divide the long utterances into 2-second chunks and calculate speaker similarity separately. Quantitative results are given in Table~\ref{tab:temporal}. First, the \textit{drift\_slope} is negative for every system: speaker similarity to the reference decays as the utterance progresses. Crucially, the gold reference also exhibits negative drift (\textit{gold\_drift\_slope}), so a portion of the decay is inherent to the chunked-similarity measurement rather than a model defect; drift is therefore best read relative to its gold counterpart. Secondly, we can observe stability with this setup. CosyVoice and VITS are the most consistent ones (low \textit{consistency\_std}, $0.021$/$0.023$; small \textit{drift\_slope}, $-0.008$/$-0.013$; and small first-to-last change), whereas Chatterbox and OmniVoice are the least stable, with the largest dispersion ($0.033$--$0.034$) and the steepest decay (\textit{first\_last\_delta} of $-0.061$ and $-0.051$ respectively). VoxCPM2 has a notable case: although its mean similarity is mid-pack, its end-to-end change is essentially zero (\textit{first\_last\_delta} $=-0.0003$) and its gold-relative drift is the smallest. It means that its voice identity stays stable over long passages. These differences matter most for long-form deployment such as audiobook or narration synthesis, where slow identity drift accumulates over minutes of audio. Figs.~\ref{fig:temporal-drift} visualize quantitative results.

\begin{table}[t]
\centering
\caption{Long-form temporal consistency. $\bar{n}_{\text{chunks}}$ = mean chunks
per utterance; sim = mean\_sim\_vs\_ref; std = consistency\_std\_mean (lower is
more stable); drift = drift\_slope\_mean; g-drift = gold\_drift\_slope\_mean;
$\Delta_{\text{fl}}$ = first\_last\_delta\_mean.}
\label{tab:temporal}
\setlength{\tabcolsep}{6pt}
\begin{tabular}{l c rrrrr}
\toprule
System & $\bar{n}_{\text{chunks}}$ & sim & std & drift & g-drift & $\Delta_{\text{fl}}$ \\
\midrule
Chatterbox (ZS)     & 12.08 & 0.8309 & 0.0325 & -0.0391 & -0.0279 & -0.0613 \\
CosyVoice           & 11.14 & 0.8533 & 0.0212 & -0.0077 & -0.0068 & -0.0094 \\
OmniVoice           & 11.73 & 0.8054 & 0.0336 & -0.0329 & -0.0340 & -0.0505 \\
OmniVoice (ZS)      & 11.88 & 0.8191 & 0.0282 & -0.0184 & -0.0110 & -0.0276 \\
VITS                & 10.42 & 0.8545 & 0.0226 & -0.0125 & -0.0207 & -0.0175 \\
VoxCPM2 (ZS)        & 11.89 & 0.8493 & 0.0233 & -0.0115 & \phantom{-}0.0049 & -0.0003 \\
\bottomrule
\end{tabular}
\end{table}


\begin{figure}[t]
\centering
\includegraphics[width=0.92\textwidth]{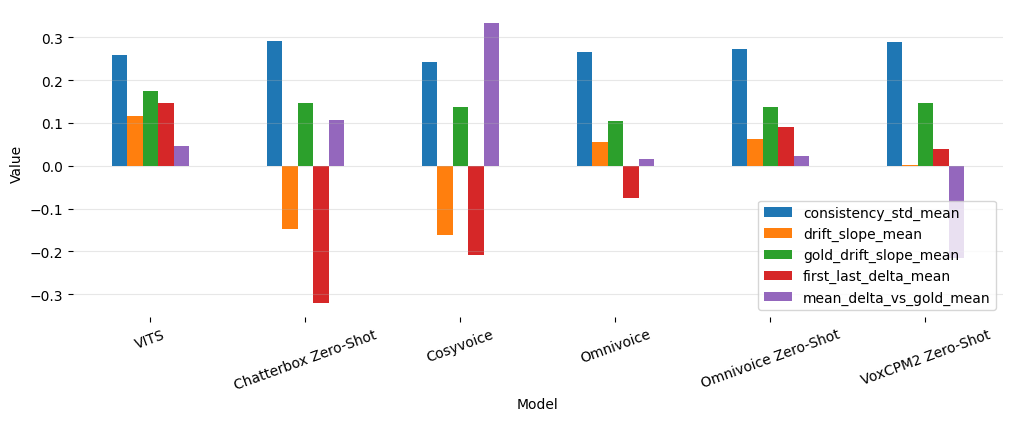}
\caption{Naturalness (UTMOS-v2) drift/consistency quantities per system:
\textit{consistency\_std\_mean}, \textit{drift\_slope\_mean},
\textit{gold\_drift\_slope\_mean}, \textit{first\_last\_delta\_mean} and
\textit{mean\_delta\_vs\_gold\_mean}. 
}
\label{fig:temporal-utmos-drift}
\end{figure}

We repeat the consistency analysis with UTMOS-v2 while we evaluate chunks separately and observe stability. Results are given in Table~\ref{tab:temporal-utmos} and Figs.~\ref{fig:temporal-utmos-drift} visualise numerical results. First, naturalness is far more volatile: the per-chunk standard deviation is roughly an order of magnitude larger ($0.23$--$0.29$) than for speaker similarity ($\sim$$0.02$--$0.03$), so quality fluctuates strongly across an utterance even when identity is stable. CosyVoice is the most stable from the perspective of naturaless. On the other hand, Chatterbox and VoxCPM2 are the least consistent models. Second, the gold reference shows a \emph{positive} UTMOS drift for every system ($+0.10$ to $+0.17$): natural speech tends to score lower on its opening chunk and rise thereafter, so the expected temporal profile is upward. Against this baseline, VITS and both OmniVoice modes drift upward like gold, whereas Chatterbox drifts \emph{downward} ($-0.149$) and shows by far the largest first-to-last drop ($-0.321$); CosyVoice drifts only mildly downward ($-0.041$). Third, the mean offset from gold (\textit{mean\_delta\_vs\_gold}) mirrors Table~\ref{tab:overall}: CosyVoice sits well above gold ($+0.31$), VoxCPM2 clearly below ($-0.22$), and the remaining systems are close to natural speech. A system can therefore be identity-stable yet naturalness-unstable: Chatterbox is unstable on both axes, while CosyVoice holds its identity but, in its phoneme mode, sheds perceived quality over the utterance (Sec.~\ref{sec:results-ablation}).

\begin{table}[t]
\centering
\caption{Long-form temporal consistency of \emph{naturalness} (UTMOS\,v2) per
system. UTMOS = mean\_utmos\_mean; std = consistency\_std\_mean; drift =
drift\_slope\_mean; g-drift = gold\_drift\_slope\_mean;
$\Delta_{\text{fl}}$ = first\_last\_delta\_mean;
$\Delta_{\text{gold}}$ = mean\_delta\_vs\_gold\_mean. Positive drift means quality
rises over the utterance; gold drifts upward for all systems.}
\label{tab:temporal-utmos}
\setlength{\tabcolsep}{6pt}
\begin{tabular}{l rrrrrr}
\toprule
System & UTMOS & std & drift & g-drift & $\Delta_{\text{fl}}$ & $\Delta_{\text{gold}}$ \\
\midrule
Chatterbox (ZS)   & 3.026 & 0.293 & -0.149 & 0.148 & -0.321 & \phantom{-}0.107 \\
CosyVoice (raw)   & 3.243 & 0.233 & -0.041 & 0.124 & -0.067 & \phantom{-}0.305 \\
OmniVoice (raw)   & 2.938 & 0.267 & \phantom{-}0.057 & 0.105 & -0.076 & \phantom{-}0.015 \\
OmniVoice (ZS)    & 2.941 & 0.272 & \phantom{-}0.063 & 0.138 & \phantom{-}0.090 & \phantom{-}0.022 \\
VITS              & 2.992 & 0.260 & \phantom{-}0.118 & 0.175 & \phantom{-}0.147 & \phantom{-}0.046 \\
VoxCPM2 (ZS)      & 2.708 & 0.288 & \phantom{-}0.001 & 0.148 & \phantom{-}0.039 & -0.216 \\
\bottomrule
\end{tabular}
\end{table}
 
 
\subsection{CosyVoice Front-End Ablation: Raw Text vs. Phonemes}
\label{sec:results-ablation}

Since the Turkish language has a largely transparent orthography (grapheme-to-phoneme conversion is close to regular), a natural question is whether an explicit phoneme front-end helps CosyVoice at all. Table~\ref{tab:cosy-frontend} isolates this by comparing the raw-text and phoneme configurations on every overall metric. The differences are small and mixed: no configuration is consistently better, and speaker identity is essentially unaffected (\textit{speaker-similarity} differs by $0.001$). The phoneme front-end reproduces the gold sibilance ratio almost exactly ($0.084$ vs.\ gold $0.084$, against $0.050$ for raw text) and is marginally ahead on the SQUIM and DNSMOS-Pro/UTMOS-style scores, whereas raw text is clearly ahead on \textit{dnsmos-pro-bvcc} ($+0.283$), \textit{natscore} ($+0.281$), and \textit{ttsds} ($+0.743$). For a shallow-orthography language such as Turkish, this near-parity is expected: phonemisation adds little that the model cannot already recover from the text. The one place the front-ends diverge meaningfully is in temporal behaviour. On the identity axis, they are nearly indistinguishable (mean similarity $0.853$ vs.\ $0.853$, \textit{consistency\_std} $0.021$ vs.\ $0.022$, \textit{first\_last\_delta} $-0.009$ vs.\ $-0.014$). On the naturalness axis, however, raw text drifts only mildly downward over an utterance (UTMOS \textit{drift\_slope} $-0.041$, \textit{first\_last\_delta} $-0.067$) while the phoneme mode decays markedly ($-0.162$ and $-0.208$), even though its mean UTMOS is fractionally higher ($3.276$ vs.\ $3.243$). In alternative terms, utilizing a phoneme-based front-end optimizes mean architectural naturalness only minimally, a benefit that is heavily counterbalanced by a pronounced degradation in intra-utterance consistency.

\begin{table}[t]
\centering
\caption{CosyVoice front-end ablation: raw text (CV-R) vs.\ phonemes (CV-P), mean
over the full test set. $\Delta_{\text{P$-$R}}$ is the phoneme$-$raw difference;
Gold is the natural reference. Differences are small for most metrics and speaker
identity is essentially unchanged.}
\label{tab:cosy-frontend}
\setlength{\tabcolsep}{6pt}
\resizebox{\textwidth}{!}{%
\begin{tabular}{l| rrr |r}
\toprule
Metric & CV-R (raw) & CV-P (phon.) & $\Delta_{\text{P$-$R}}$ & Gold \\
\midrule
audiobox-pq            & 7.793 & 7.757 & $-0.036$ & 7.657 \\
brouhaha                & 60.249 & 58.140 & $-2.109$ & 55.309 \\
dnsmos\_pro             & 3.770 & 3.853 & $+0.083$ & 3.601 \\
dnsmos-pro-bvcc       & 3.643 & 3.360 & $-0.283$ & 3.698 \\
dnsmos-pro-nisqa      & 3.770 & 3.853 & $+0.083$ & 3.601 \\
dnsmos-pro-vcc2018    & 2.925 & 2.928 & $+0.003$ & 3.171 \\
speaker-similarity     & 0.878 & 0.877 & $-0.001$ & 0.894 \\
squim-pesq             & 0.993 & 0.993 & \phantom{$+$}$0.000$ & 0.992 \\
squim-si-sdr          & 25.736 & 26.215 & $+0.479$ & 24.455 \\
squim-stoi             & 3.753 & 3.818 & $+0.065$ & 3.639 \\
utmos-v2               & 3.441 & 3.501 & $+0.060$ & 3.079 \\
low-level-lufs        & -22.443 & -24.376 & $-1.933$ & -25.151 \\
low-level-sib.-ratio & 0.050 & 0.084 & $+0.034$ & 0.084 \\
scoreq-synthetic-nr   & 4.063 & 4.082 & $+0.019$ & 3.782 \\
speechlmscore           & 1.672 & 1.708 & $+0.036$ & 1.703 \\
whisqa                  & 4.720 & 4.655 & $-0.065$ & 4.331 \\
natscore                & -1.233 & -1.514 & $-0.281$ & -1.384 \\
ttsds                   & 79.735 & 78.992 & $-0.743$ & 77.864 \\
\bottomrule
\end{tabular}}
\end{table}

\section{Discussion}
\label{sec:discussion}

\subsection{No Single Winner Across Metric Families}
The primary result of this benchmark is that the best TTS system for Turkish is undefined without fixing target objective. As illustrated by the per-metric performance in Table~\ref{tab:overall}, the optimal architectures partition almost perfectly along specific metric classification families. VITS demonstrates superior efficacy within the DNSMOS-Pro domain, whereas OmniVoice consistently outperforms other models across the SQUIM intelligibility and signal metrics. Concurrently, CosyVoice exhibits robust performance in learned-naturalness and distributional evaluations (UTMOS-v2, SCOREQ, WhisQA, TTSDS), while VoxCPM2 excels in perceptual-naturalness and speaker-fidelity metrics (NatScore, speaker similarity, Brouhaha). This clustered result is more informative than a regular ranking, since a user might pick one of the systems for a specific goal: OmniVoice~ZS or VITS, where a clean, intelligible signal is paramount; CosyVoice or VoxCPM2, where perceived naturalness and faithful voice cloning matter. 

\subsection{Naturalness Predictors vs. Intelligibility/Signal Metrics}
The two metric groups are not merely different. Several systems beat the gold reference on signal-oriented predictors (VITS on DNSMOS, OmniVoice~ZS on SQUIM SI-SDR, and AudioBox-PQ), which is implausible for genuine naturalness and instead reveals that these predictors reward clean, low-noise, smoothly voiced audio. A conventional model such as VITS, whose vocoder produces very tidy spectra, can therefore top DNSMOS while sounding less natural than an expressive zero-shot system. The learned-naturalness and distributional metrics (UTMOS-v2, WhisQA, TTSDS) align better with perceptual realism and rank the expressive LLM/diffusion systems (chiefly CosyVoice) highest. We therefore treat the signal-clean metrics as quality floors (is the audio clean and intelligible?) and the naturalness/distributional metrics as the more decision-relevant indicators of human-likeness, and we caution that optimising the former can be mildly anti-correlated with the latter.

\subsection{Sensitivity to Utterance Length}
Utterance length is a first-order factor that an averaged leaderboard obscures. All systems degrade on short utterances, but the degree varies enormously: the CosyVoice/OmniVoice/VoxCPM2 families stay relatively flat, while Chatterbox collapses across nearly every metric on short input and recovers only on medium/long utterances. Because the gold reference also loses speaker-similarity on short audio, length-stratified reporting is needed both to surface genuine model weaknesses (Chatterbox short-form) and to avoid mistaking a short-audio measurement artifact (speaker-embedding instability) for a synthesis defect. Deployments dominated by short prompts (voice assistants, IVR, command responses) should weight the short-group results far more heavily than the overall averages.

\subsection{Stability in Long-Form Synthesis}
The temporal analysis adds a dimension that single-utterance metrics cannot capture: how steady a system remains across a long passage, along two distinct axes: speaker identity and perceived naturalness. On the identity axis, similarity to the reference drifts downward for every system, but only relative to the gold drift is this meaningful. On that basis, CosyVoice and VITS are the most stable, Chatterbox and OmniVoice the least, and VoxCPM2's near-zero first-to-last change makes it attractive for minutes-long content where small per-chunk drifts would otherwise compound. The \emph{naturalness} axis behaves differently and is arguably the more revealing of the two: per-chunk quality fluctuates an order of magnitude more than identity, and its expected profile is upward, since gold UTMOS rises across chunks for every system. VITS and both OmniVoice modes follow this profile, but Chatterbox drifts downward and shows the largest first-to-last drop; the same downward pattern is even stronger for CosyVoice's phoneme front-end (Sec.~\ref{sec:results-ablation}). The key lesson is that identity stability does not imply naturalness stability: a system can hold its voice yet shed perceived quality over the utterance, whereas a per-utterance average would simply call it \textit{good and stable}. For audiobook, dubbing, and narration use-cases, where these effects accumulate over minutes, long-form deployment should therefore monitor both identity drift and naturalness drift rather than trusting a single per-utterance score.

\subsection{Distance to Natural Speech (Gold) and Metric Reliability}
Anchoring every metric to gold doubles as a sanity check on the metrics themselves. Where synthetic systems surpass gold (DNSMOS, SQUIM, AudioBox-PQ, NatScore, parts of TTSDS), the metric is rewarding something other than human-likeness and should be interpreted with care for Turkish. The one dimension on which gold reliably tops all systems is speaker similarity, which behaves as expected and is thus the most trustworthy single indicator of cloning fidelity in our battery. We recommend reporting all objective scores alongside the gold value so that \textit{better than real speech} artifacts are visible rather than hidden inside a normalised aggregate. The temporal view sharpens this point: CosyVoice not only exceeds gold UTMOS on average but especially in its phoneme mode (Sec.~\ref{sec:results-ablation}) follows a different within-utterance trajectory (declining where natural speech rises), while VoxCPM2 sits below gold throughout a structure that any single averaged score conceals.

\section{Conclusion}
\label{sec:conclusion}

We present a multi-metric benchmark of Turkish text-to-speech with various objective speech quality assessment metrics. We conduct experiments with four modern TTS models (both finetuned and zero-shot configurations) and one conventional TTS model on a targeted Turkish testset. Gold records of the speaker are added to the results for analyzing the shift between real and synthetic speech. We observe evaluations on several objective metrics, investigate utterance-length breakdown, and a long-form temporal-consistency analysis. Performance is not concentrated in any single system; instead, leadership is distributed across metric families, with each system excelling within a distinct group: VITS on DNSMOS-Pro metrics, OmniVoice~(ZS) on SQUIM intelligibility and signal-quality metrics, CosyVoice on learned-naturalness and distributional metrics, and VoxCPM2 on perceptual-naturalness and speaker-fidelity metrics. Several systems surpass gold on signal-clean predictors, exposing a tension between metrics that reward clean, intelligible audio and those that track human-likeness; this dissociation motivates a panel of complementary metrics anchored to gold rather than a single aggregate. The length and temporal analyses further show that short utterances are the hardest setting, eliciting a pronounced collapse in Chatterbox, whereas CosyVoice and VITS are the most temporally stable, and VoxCPM2 shows essentially no first-to-last identity drift. Speaker similarity emerges as the most trustworthy single indicator of cloning fidelity. This study is restricted to objective metrics and relies on a single Turkish test set with limited speaker coverage, evaluated using general-purpose multilingual checkpoints; absolute scores should therefore be interpreted relative to the gold anchor rather than in isolation. Future work should incorporate human evaluation through MOS and preference studies, extend the analysis to a broader range of speakers and domains, and investigate long-form stability over multi-minute passages.

\bibliographystyle{splncs04}
\bibliography{references}

\end{document}